\documentclass[12pt]{article}
\input{epsf.sty}
\usepackage{graphics}

\def\mydate{24 December 2001}
\def\ignore#1{{}}

\let\oldtheequation=\theequation
\def\doteqs#1{\setcounter{equation}{0}
            \def\theequation{{#1}.\oldtheequation}}
\newcounter{sxn}
\def\sx#1{\addtocounter{sxn}{1} \vskip 1.cm  \goodbreak
\noindent{\large\bf\leftline{\thesxn.~~#1}} \nobreak \vskip -.5cm}
\def\sxn#1{\sx{#1} \doteqs{\thesxn}}

\newcounter{axn}

\date{}

\newdimen\mybaselineskip
\renewcommand{\baselinestretch}{1.25}

\newcommand{\beeq}{\begin{equation}}
\newcommand{\eneq}{\end{equation}}
\newcommand{\beqn}{\begin{eqnarray}}
\newcommand{\eeqn}{\end{eqnarray}}

\def\mybig{\displaystyle \strut }

\def\dd{\partial}
\def\la{\raise.16ex\hbox{$\langle$}\lower.16ex\hbox{}  }
\def\ra{\, \raise.16ex\hbox{$\rangle$}\lower.16ex\hbox{} }
\def\go{\rightarrow}

\def\next{{~,~~~}}
\def\onehalf{ \hbox{${1\over 2}$} }

\def\eff{{\rm eff}}
\def\P{{\rm P}}

\def\vphi{\varphi}
\def\ep{\epsilon}

\def\myfrac#1#2{{\mybig #1\over \mybig #2}}

\begin{document}

\thispagestyle{empty}

\baselineskip=14pt

{\small \noindent \mydate\hfill  OU - 397 /2001}
\rightline{\small NUC-MINN-01/17-T}


\baselineskip=40pt plus 1pt minus 1pt

\vskip 2.5cm

\begin{center}

{\Huge \bf Cosmic Shells}\\

\vspace{3.0cm}
\baselineskip=20pt plus 1pt minus 1pt

{\bf  Yutaka Hosotani and Takayuki Nakajima}\\
\vspace{.1cm}
{\it Department of Physics, Osaka University, 
Toyonaka, Osaka 560-0043, Japan}\\ 
\vspace{1.5cm}
{\bf  Ramin  G.\ Daghigh   and Joseph I.\ Kapusta}\\
\vspace{.1cm}
{\it School of Physics and Astronomy, University of
Minnesota, Minneapolis, MN 55455, USA}\\ 
\end{center}

\vskip 3.cm
\baselineskip=20pt plus 1pt minus 1pt

\begin{abstract}
When a potential for a scalar field has two local minima there arise
spherical shell-type solutions of the classical field equations
due to gravitational attraction. We establish such solutions numerically
in a space which is asymptotically de Sitter.  It generically
arises when the energy scale characterizing the scalar field potential
is much less than the Planck scale. It is shown that the mirror image of
the shell appears in the other half of the Penrose diagram.  The configuration 
is smooth everywhere with no physical singularity.
\end{abstract}

\newpage


\newpage

\sxn{Introduction}

Gravitational interactions, which are inherently attractive for ordinary
matter, can produce soliton-like objects even when such things
are strictly forbidden in flat space.  They are possible as
a consequence of the balance between repulsive and attractive forces.
One such example is a monopole or dyon solution in the pure
Einstein-Yang-Mills theory in the asymptotically anti-de Sitter
space \cite{Hosotani1, Radu, Galtsov0}.  In the pure Yang-Mills theory in
flat space  there can be no static solution at all \cite{Deser} but once
gravitational interaction is included there arise particle-like solutions
\cite{Bartnik}. Whereas all solutions are unstable in the asymptotically
flat or de Sitter space, there appear a continuum of stable monopole and
dyon solutions in the asymptotically anti-de Sitter space.  The stable
solutions are cosmological in nature; their size is typically of order
$|\Lambda|^{-1/2}$ where $\Lambda$ is the cosmological constant.

The possibility of  false vacuum black holes has also been
explored. Suppose that the potential in a scalar field
theory has two minima, one corresponding to the  true vacuum and the other
to the false vacuum.   If the universe is in the false vacuum, a bubble
of the true vacuum is created by quantum tunneling which expands with
accelerated velocity.  The configuration is called a bounce \cite{Coleman}.  
Now flip the configuration \cite{Kapusta}.  The universe is in the
true vacuum with potential $V=0$ and the inside of a sphere is excited to the 
false vacuum with $V>0$. Is such a de Sitter lump in Minkowski space possible?  
If the lump is too small it would be totally unstable.  The energy localized
inside the lump can dissipate to spatial infinity.  If the lump is big
enough the Schwarzschild radius becomes larger than the lump radius so that
the lump is inside a black hole.  The energy cannot escape to infinity.
It looks like a soliton in Minkowski space.  However, as a black hole it is a 
dynamical object.  The configuration is essentially time-dependent.
This false vacuum black hole configuration, however, does not solve
the equations at the horizon.  It has been recently proven that there
can be no such everywhere-regular black hole solution \cite{Galtsov,Bronnikov}.
Rather, false vacuum lumps in flat space evolve dynamically 
\cite{Maeda,Sato,Guth}.

The purpose of this paper is to report new solutions to the coupled
equations of  gravity and scalar field theory which display spherically
symmetric shell  structure \cite{Hosotani2}.  We demonstrate that such
structure appears when the potential for a scalar field has two local minima
and the space is asymptotically de Sitter.  In the examples we present, both
the inside and outside of the shells are de Sitter space with the same
cosmological constant.   The structure becomes possible only when the energy
scale of the scalar field potential becomes small compared with the Planck 
scale.  While such shell structures might not be easy to create in the present 
universe,  it is quite plausible that they could have been created during a
phase  transition early in the universe \cite{Maeda}.
A similar configuration has been investigated in ref.\ \cite{Dymnikova}.

The plan of our paper is as follows.  In section 2 we precisely state the 
problem, solve the field equations in those regions of space-time where they can 
be linearized, and sketch the solution in the nonlinear shell region in static 
coordinates.  In section 3 we solve the nonlinear equations in the shell region 
and display the dependence on the parameters of the theory.  In section 4 we 
extend the solution from static coordinates, which have a coordinate 
singularity, to other coordinate systems that do not, thereby displaying the 
existence and character of the solution throughout the full space-time manifold.  
In section 5 we study the stability of the classical solution to quantum 
fluctuations. Our summary and conclusions are given in section 6.

\sxn{Shell Solutions in Static Coordinates}

\begin{figure}[tbh]
\centering \leavevmode 
\rotatebox{90}{
\epsfxsize=7.5cm \epsfbox{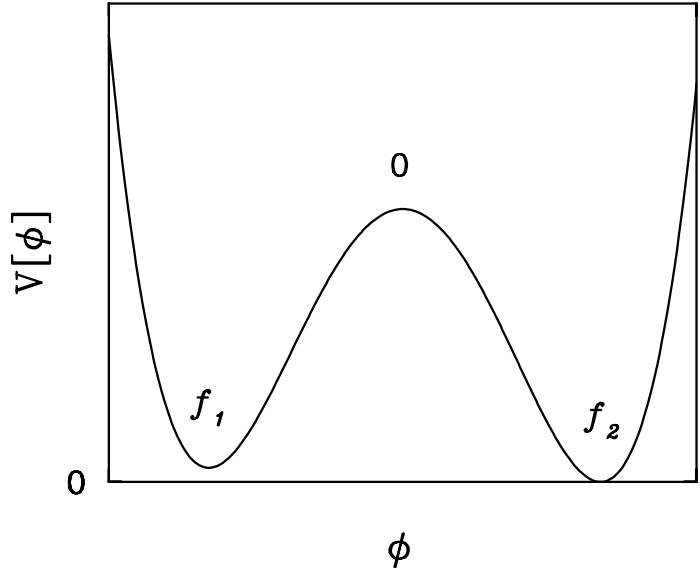} }
\caption{The potential $V[\phi]$.}
\label{potential}
\end{figure}

Consider a scalar field coupled to gravity with the Lagrangian 
\beeq
{\cal L} = 
{1\over 16\pi G} ~ R
   + {1\over 2}  \phi_{;\mu} \phi^{;\mu}   - V[\phi] 
\eneq
where the potential $V[\phi]$ has
two minima at $f_1$ and $f_2$ separated by a barrier. See fig.\ 
\ref{potential}.

We look for spherically symmetric configurations in which 
the metric of space-time is written as
\beeq
ds^2 = 
-\frac{H}{p^2}dt^2 + \frac{dr^2}{H} 
  +r^2 (d\theta^2  + \sin^2 \theta d\vphi^2) ~.
\label{ourmetric1}
\eneq
The functions $\phi$, $H$, and $p$ depend only on $r$ and $t$.
A tetrad basis is chosen, in a region $H>0$, as
\beeq
e_0 = \frac{\sqrt{H}}{p} \, dt \next
e_1 = \frac{1}{\sqrt{H}} \, dr\next
e_2 = r d\theta \next e_3 = r\sin\theta d\vphi ~.
\label{tetrad1}
\eneq
The components of the energy-momentum tensor in the tetrad basis, 
$T_{ab} = {e_a}^\mu {e_b}^\nu T_{\mu\nu}$, are
\beqn
&&\hskip -.8cm
T_{00} = {1\over 2} \bigg( \frac{p^2}{H} \, \dot\phi^2 
  + H \phi'^2 \bigg) + V[\phi] \cr
&&\hskip -.8cm
T_{11} = {1\over 2} \bigg( \frac{p^2}{H} \, \dot\phi^2 
  + H \phi'^2 \bigg) - V[\phi] \cr
&&\hskip -.8cm
T_{22} = T_{33}  
={1\over 2} \bigg( \frac{p^2}{H} \, \dot\phi^2 
   - H \phi'^2 \bigg) - V[\phi] \cr
\noalign{\kern 5pt}
&&\hskip -.8cm
T_{01} = -p \dot\phi \phi'  ~.
\label{tensor1}
\eeqn
Here dot and prime indicate $t$- and $r$-derivatives, respectively.

The scalar field satisfies
\beeq
p \frac{\dd}{\dd t} \bigg( \frac{p}{H} \dot\phi \bigg)
- {p\over r^2} {\dd\over \dd r} \bigg( {r^2 H\over p} \phi' \bigg)
+ V'[\phi] = 0 ~.
\label{scalar1}
\eneq
We introduce the integrated mass function $M(t,r)$ by
\beeq
H = 1 - {2GM\over r} ~.
\label{mass1}
\eneq
The Einstein equations are 
\beqn
&&\hskip -.8cm
M = \int_0^r 4\pi r^2 dr \, T_{00} ~, 
\label{Ein1} \\
&&\hskip -.8cm
{p'\over p} = - 4\pi G r \bigg\{ {p^2\over H^2} \, \dot\phi^2 + 
  \phi'^2 \bigg\} ~,  
\label{Ein2} \\
&&\hskip -.8cm
{\dot H\over H} = - 8\pi G r \dot\phi \phi' ~,
\label{Ein4}\\
&&\hskip -.8cm
{p\over 2} \bigg\{ {\dd\over \dd t} \bigg( {p\dot H\over H^2} \bigg) 
+ {\dd\over \dd r} \bigg( {H'\over p} - {2Hp'\over p^2} \bigg) \bigg\}
+ {1-H\over r^2} 
= 4\pi G\bigg( {p^2\over H} \,\dot\phi^2 - H \phi'^2 \bigg) ~.
\label{Ein3}
\eeqn
One of the equations, Eq.\ (\ref{Ein3}), is redundant as it follows from Eqs.\
(\ref{scalar1}), (\ref{Ein1}), (\ref{Ein2}), and (\ref{Ein4}).

We shall seek static solutions for which the set of equations 
reduces to
\beqn
&&\hskip -.8cm
\phi''(r) + \Gamma_\eff (r)  \phi'(r) = {1\over H} ~ V'[\phi] ~,
\label{scalar2}  \\
&&\hskip -.8cm
M(r) =\int_0^r 4\pi r^2 dr \, \left\{ {1\over 2} H \phi'^2 
   +V[\phi] \right\} ~,
\label{mass2}
\eeqn
where
\beqn
\Gamma_\eff \equiv {2\over r} + 4\pi G r \phi'^2 + {H'\over H} ~.
\eeqn 
Eq.\ (\ref{scalar2}) can be interpreted as an equation for a particle with
a coordinate $\phi$ and time $r$.  Except for a factor $1/H$ this particle
moves in a potential $U[\phi]=-V[\phi]$.  The coefficient $\Gamma_\eff(r)$
represents time ($r$)-dependent friction.  

The potential $V[\phi]$ is supposed to have two minima, at $f_1$ and $f_2$.  
We are looking for a solution  which starts at $\phi \sim f_1$, moves close
to $f_2$, and comes back to $f_1$ at $r=\infty$.  The particle's
potential $U[\phi]$ has two maxima.  The particle begins near the top
of one hill, rolls down into the valley and up the other hill, turns
around and rolls down and then back up to the top of the original hill.
This is impossible in flat space, as $\Gamma_\eff$ is positive-definite
so that the particle's energy dissipates and it cannot climb back
to its starting point.

In the presence of gravity the situation changes.  The non-vanishing
energy density can make $H$ a decreasing function of $r$ so that
$\Gamma_\eff$ becomes negative.  The energy lost by the particle during the 
initial rolling down can be regained on the return path by negative
friction, or thrust.  Indeed, this happens.

Let us set up the problem more precisely.  We take a quartic potential
with $f_1 <0 < f_2$;
\beqn
&&\hskip -.8cm
V'[\phi] = \lambda \phi(\phi- f_1)(\phi-f_2) ~,   \cr
\noalign{\kern 5pt}
&&\hskip -.8cm
V[\phi] = \frac{\lambda}{4}(\phi-f_2) 
 \bigg\{ \phi^3 - {1\over 3} (f_2 + 4f_1) \phi^2 
  -{1\over 3} f_2 (f_2- 2f_1 )  (\phi+f_2) \bigg\} ~. 
\label{potential1}
\eeqn
Here $V[f_2] = 0$, and $V[0]$ is a local maximum for
the barrier separating the two minima.
Define $f=(|f_1| + f_2)/2$ and $\Delta f = f_2 - |f_1|$.   In case
$\Delta f > 0$, $\phi=f_1$ corresponds to a false vacuum with the 
energy density $\ep = V[f_1] = {2\over 3} \lambda f^3 \Delta f >0$,
whereas $\phi=f_2$ corresponds to a true vacuum with a vanishing  energy
density. As we shall discuss in detail below, the positivity of the energy
density $\ep$ plays an important role for the presence of shell structure, but
the vanishing $V[f_2]$ is not essential as we see below.   In a more
general potential  it could be that $V[f_2] > V[f_1]$.

We look for solutions with $\phi$ starting at the origin $r=0$ very close to
$f_1$. There is only one parameter to adjust: $\phi_0 \equiv \phi(0)$. 
The  behavior of a solution near the origin is given by
\beqn
&&\hskip -1.cm
\phi = \phi_0 + \phi_2 r^2 + \cdots ~, 
\hskip 2cm \phi_2 = {1\over 6} \, V'[\phi_0] ~,\cr
&&\hskip -1.cm
p = 1 + p_4 r^4 + \cdots ~, 
\hskip 2.3cm p_4 = - 4\pi G \phi_2^2 ~,  \cr
&&\hskip -1.cm
M = m_3 r^3 + \cdots, 
\hskip 2.7cm  m_3 = {4\pi\over 3} \, V[\phi_0] ~, \cr
&&\hskip -1.cm
H = 1 - 2 G m_3 r^2 + \cdots ~.
\label{origin1}
\eeqn
Given $\phi_0$ the equations determine the behavior of a configuration
uniquely.  For most values of $\phi_0$ the corresponding configurations
are unacceptable.   As $r$ increases, $\phi(r)$ either approaches
0 (the local maximum of $V[\phi]$) or goes to $ - \infty$.
Other than the two trivial solutions, corresponding to the false and true
vacua, we have found a new type of solution.

There are four parameters in the model, one of which, the gravitational
constant $G= m_{\rm P}^{-2}$, sets the scale.   The other three are
$\lambda$, $f_1$, and $f_2$ or, equivalently, the three dimensionless
quantities $f/m_{\rm P}$, $\Delta f/f$, and $\lambda$.  We have  
explored only a limited region in the parameter space.  The moduli 
space of solutions depends critically on $f/m_{\rm P}$ and  $\Delta f/f$,
but seems to depend little on $\lambda$.    Nontrivial solutions appear
as 
$f/m_{\rm P}$ becomes small.

If $\phi_0 < f_1$, $\phi(r)$ monotonically
decreases as the radius increases to diverge to $-\infty$.
On the way $H(r)$ crosses zero.
Suppose instead that $f_1 < \phi_0 <0$ and $\phi_0$ is not too close to
$f_1$.  In the particle analogue, the particle starts to roll down the 
hill under the action of $U[\phi]= - V[\phi]$.  It approaches $\phi=0$, and 
oscillates around it.  In the meantime $H(r)$ crosses zero.  

\begin{figure}[tbh]
\centering \leavevmode 
\rotatebox{-90}{
\epsfxsize=8.cm \epsfbox{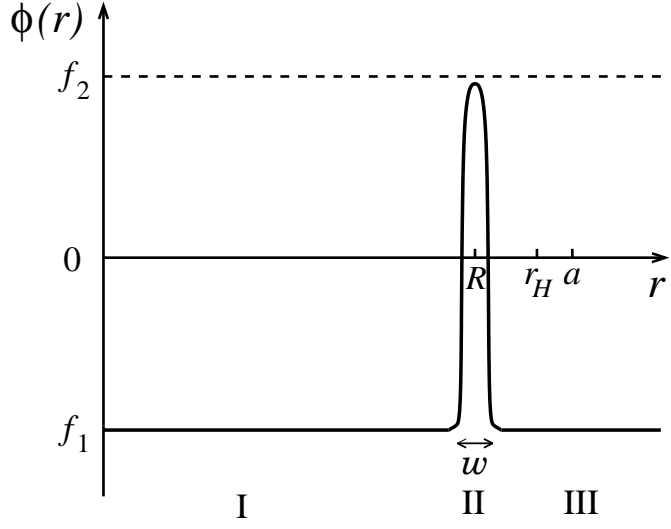}}
\caption{Schematic behavior of $\phi(r)$.}
\label{global-phi}
\end{figure}

\begin{figure}[tbh]
\centering \leavevmode 
\rotatebox{-90}{
\epsfxsize=8.cm \epsfbox{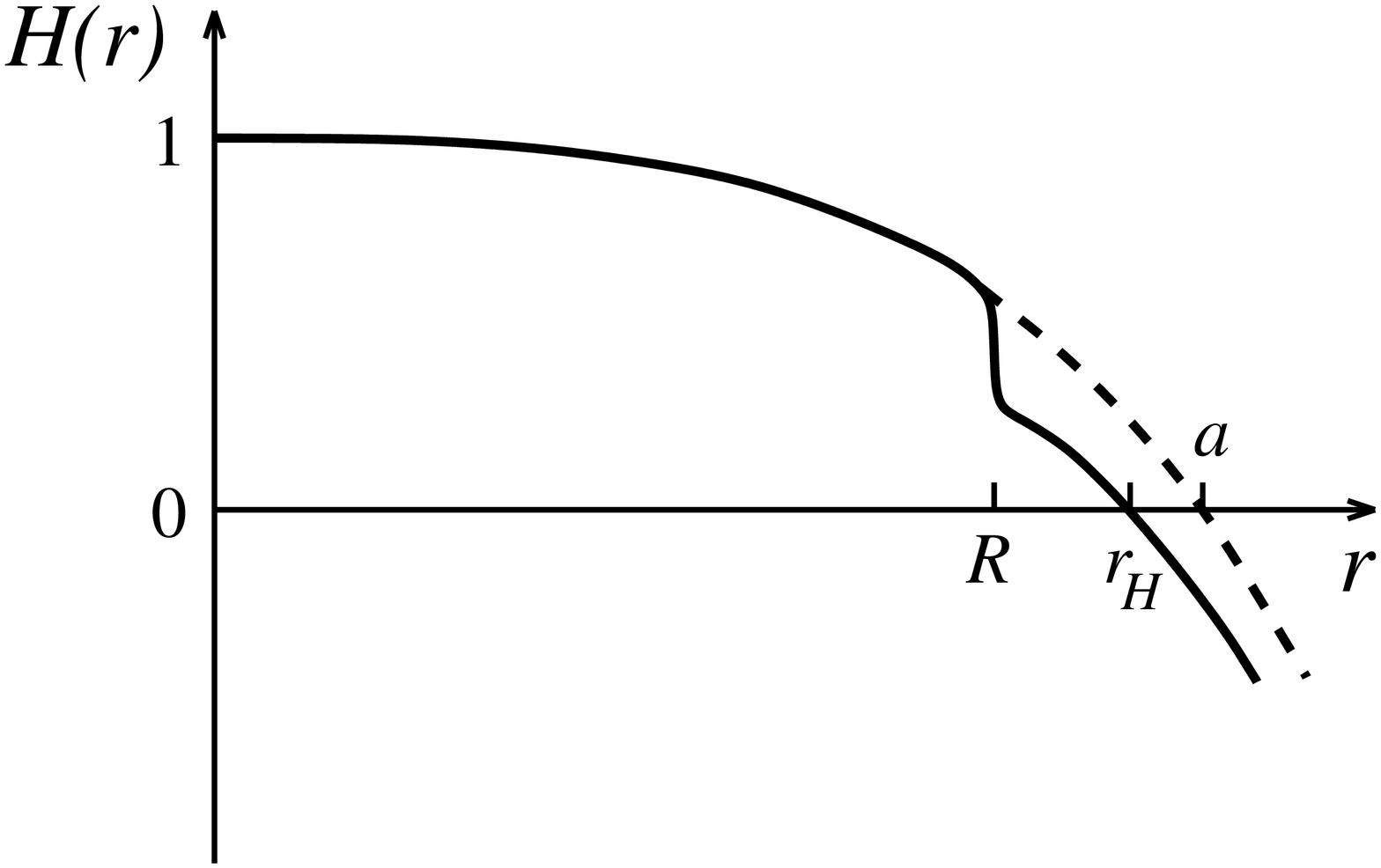}}
\caption{Schematic behavior of  $H(r)$.}
\label{global-H}
\end{figure}

Now suppose that $\phi_0$ is very close to, but still greater than,
$f_1$: $\phi(0) = f_1 + \delta \phi(0)$ with 
$0 < \delta\phi(0)/f \ll 1$.  A schematic of the resulting solution is 
displayed in fig.\ \ref{global-phi} and fig.\ \ref{global-H}.  We divide
space into three regions in the  static coordinates: region I $( 0 \le r <
R_1)$, region II $(R_1 < r < R_2)$, and region III $(R_2 < r)$.  It turns
out that
$\phi(r)$ varies little from 
$f_1$ in regions I and III so that the equation of motion for $\phi$ may be 
linearized in those regions.  In region II the field deviates strongly and the 
full set of nonlinear equations must be solved numerically.  This is the region 
in which we shall find shell structure.  $H(r)$ deviates from the de
Sitter value significantly.  In region III  the spacetime is approximately
de Sitter again.  $H(r)$ crosses zero at $r_H$.

In region I the space-time is approximately de Sitter.
\beeq
T_{00} = \ep \next
H = 1 - {r^2\over a^2} \next a = \sqrt{ {3\over 8\pi G \ep}} \next
p=1 ~.
\label{metric2a}
\eneq
The equation for $\phi(r)$ can be linearized with
$\phi(r) = f_1 + \delta\phi(r)$.  In terms of $z \equiv r^2/a^2$,
\beqn
\Bigg\{ z(1-z) {d^2\over dz^2} 
+ \bigg( {3\over 2} - {5\over 2} z \bigg) {d\over dz}
- {1\over 4} \omega^2 a^2 \Bigg\} \, \delta\phi = 0 ~,
\label{scalar3}
\eeqn
where $\omega^2 = V''[f_1]$.  This is Gauss' hypergeometric equation.
The solution which is regular at $r=0$ is
\beqn
\delta \phi(r) = \delta \phi (0) \cdot
F( \hbox{$\frac{3}{4}$} + i \kappa , \hbox{$\frac{3}{4}$} - i \kappa ,
  \hbox{$\frac{3}{2}$} ; z) ~,
\label{phi1}
\eeqn
where
\beqn
&&\hskip -1.cm
\kappa = \onehalf \sqrt{ \omega^2 a^2 - \hbox{${9\over 4}$} }  \next
\omega^2 a^2 ={9 m_\P^2 \over 8 \pi f \Delta f} 
    \bigg( 1 - {\Delta f\over 2f} \bigg) ~.
\eeqn
We shall soon see that a solution with shell structure appears for
$\omega a \gg 1$ with a particular choice of $\delta \phi(0)$.
The ratio of $\delta \phi'(r)$ to $\delta \phi(r)$ is given by
\beeq
{\delta \phi'(r) \over \delta \phi(r)}
= {4r\over 3 a^2} \Big( \kappa^2 + {9\over 16} \Big)
{F( \hbox{$\frac{7}{4}$} + i \kappa , \hbox{$\frac{7}{4}$} - i \kappa ,
  \hbox{$\frac{5}{2}$} ; z)  \over 
F( \hbox{$\frac{3}{4}$} + i \kappa , \hbox{$\frac{3}{4}$} - i \kappa ,
  \hbox{$\frac{3}{2}$} ; z) } 
\equiv {2r\over a^2} ~ J(z)~.
\label{phi2}
\eneq

The deviation from $f_1$ at the origin, $\delta\phi(0)$, needs to be
very small for an acceptable solution.  The behavior of the
hypergeometric function for $\kappa \gg 1$ and $0<z<1$ is given by 
\cite{Bateman}
\beqn
&&\hskip -1.cm
F(a+ i\kappa, a - i\kappa, c; z) 
\sim {\Gamma(c)\over 2 \sqrt{\pi} } ~ 
\kappa^{{1\over 2} - c} ~
z^{- {c\over 2} + {1\over 4} } (1-z)^{{c\over 2} - {1\over 4} -a } ~
 \exp \left\{ 2\kappa \sin^{-1} \sqrt{z} \right\} ~.
\label{geometricF}
\eeqn
The ratio $\delta\phi(r)/\delta \phi(0)$ grows exponentially as $r$ increases
like $(4\kappa)^{-1} z^{-1/2} (1-z)^{-1/4} \, 
\exp \left\{ 2\kappa \sin^{-1} \sqrt{z} \right\}$.
At the end of region I, $\delta\phi/|f_1|$ needs
to be very small for the linearization to be valid.  The ratio 
of $F'(z)$ to $F(z)$, $J(z)$ in
(\ref{phi2}),  is given by
\beeq
J(z) = {\kappa \over \sqrt{ z(1-z) } } \hskip 1cm
 \hbox{for } \kappa \gg 1 ~,~ 0<z<1~.
\label{geometricF2}
\eneq

In region II, $\phi(r)$ varies substantially and
the nonlinearity of  the equations plays an essential role.
In this region the equations must be solved numerically.  With fine
tuning of the value of $\delta\phi(R_1)$ nontrivial shell solutions
will be found.

The algorithm is the following.  First $\delta\phi(R_1)$ is chosen and
$\delta\phi'(R_1)$ is evaluated by (\ref{phi2}) and (\ref{geometricF2}).
To the order in which we work the metric is
$H(R_1) = 1 - (R_1/a)^2$ and $p(R_1)=1$.  With these boundary conditions
Eqs.\ (\ref{scalar2}) and (\ref{mass2}) are numerically solved.

The  behavior of solutions in region II is displayed in fig.\ 4.
When the specific values of the input parameters are chosen to be
$\lambda=0.01$, $f/m_\P = 0.002$, and $\Delta f/f = 0.002$, then the output
parameters are: $a/l_\P=2.365 \times 10^7$, $\kappa=3.344 \times 10^3$,
$R_1/ l_\P = 1.76104695 \times 10^7$ and $\delta\phi(R_1)/m_\P=10^{-5}$. 
(Here $l_\P$ is the Planck length.)  The field
$\phi(r)$ approaches $f_1$ for $r > R_2$.
In the numerical integration $\delta\phi(R_1)$ is kept fixed while
$R_1$ is varied.  Fine-tuning to the ninth digit is necessary!
If $R_1$ is taken to be slightly bigger then $\phi(r)$ starts to deviate
from $f_1$ in the negative direction, diverging to $-\infty$ as $r$
increases.  If $R_1$ is taken to be slightly smaller then $\phi(r)$ starts 
to deviate from $f_1$ in the positive direction, heading for $f_2$ as
$r$ increases.  With just the right value the space-time becomes nearly de 
Sitter outside the shell.  The value of $\delta\phi$ at the origin ($r=0$) is 
found from (\ref{phi1}) to be $1.40 \times 10^{-2440}$, which explains why one 
cannot numerically integrate $\phi(r)$ starting from $r=0$.

The behavior of the solution in region III is easily inferred.
  From the numerical integration in region II both $H_2=H(R_2)$
and $p_2 = p(R_2)$ are determined.  In region III the metric can be
written in the form 
\beqn
&&\hskip -1.cm
H(r) = 1 - {2 G \tilde M\over r} - {r^2\over a^2} \cr
&&\hskip -1.cm
p(r) = p_2 ~.
\label{metric2}
\eeqn
Here $\tilde M$ is the mass ascribable to the shell.  Once $T_{00}$
vanishes $H(r)$ must take this form.  The value of $\tilde M$ may be
determined numerically by fitting Eq.\ (\ref{metric2}) just outside
the shell.  
The location of the horizon, $r_H$, is determined by $H(r_H) = 0$.
The field $\phi(r)$ remains very close to $f_1$ in region III. 

At this point due caution is required to continue the solution because 
the static coordinates  defined in (\ref{ourmetric1}) do not cover the
whole of space-time.  The global structure of the space-time and of the
solutions is  worked out in Section 4 where different coordinates are
employed.

The solutions we found have shell structure at a radius $R$ whose width $w$ is 
very small.  The radius $R$ is smaller than the horizon radius $r_H$, but is of 
the same order as $a$.  In the next section we shall see that $R$ becomes 
smaller as $f$ becomes smaller, but remains of order $a$ even in the $f \go 0$ 
limit. This implies that the shell structure is cosmic in size.  

An estimate of the order of magnitude of $w$ as well as the conditions necessary
for the existence of the cosmic shell are obtained by a simple
argument.  Return to Eq.\ (\ref{scalar2}).
\beeq
H \phi'' + \Big( {2H\over r} + H' \Big) \phi'
+ 4\pi G r H \phi'^3 = V'[\phi]
\label{scalar4}
\eneq
In the shell region (region II, $r\sim R$) of typical solutions, $H$
is of order one and drops sharply.   In (\ref{scalar4}) the $2H/r$ term 
is negligible compared with the $H'$ term.  The remaining terms, $H\phi''$, 
$H'\phi'$, and $V'[\phi]$ are all of the same order of magnitude
so that $f/w^2 \sim \lambda f^3$ or
\beeq
w \sim {1\over \sqrt{\lambda} f} ~.
\label{estimate1}
\eneq
Hence the thickness of the shell is determined by the parameters
in the scalar field potential.  The radius $R$ is smaller than but of the 
same order as $a$.  The solutions exist only if $H'$, which is 
negative, dominates over $4\pi G r H \phi'^2$.  In other words
$w^{-1} > 4\pi G R(f/w)^2$.  Making use of $R\sim a$,  
$a^2 = 9/(16 \pi G \lambda f^3 \Delta f)$ and (\ref{estimate1}), one finds
\beqn
&&\hskip -1cm 
4\pi \bigg( {f \over m_{\rm P}} \bigg)^2 <
{w\over a} \sim {4\sqrt{\pi}\over 3} {\sqrt{f \Delta f}\over m_{\rm P}} 
\ll 1 \cr
\noalign{\kern 10pt}&&\hskip -1cm 
{\Delta f\over f} > 3 \sqrt{\pi} \bigg( {f \over m_{\rm P}} \bigg)^2 ~.
\label{estimate2}
\eeqn
We shall see in the next section that these relations are satisfied in the 
solutions obtained numerically.

\sxn{Numerical Analysis of the Nonlinear Regime}

In the last section we solved the linearized field equation for $\phi(r)$ in 
regions I and III where the deviation from $f_1$ is small, and we sketched the 
behavior in the nonlinear region II where the shell structure appears.
In this section we present numerical results for region II. As discussed in the 
last section the boundary between regions I and II,
located at the matching radius $R_1$, is rather arbitrary,
subject only to the condition that the linearization is accurate 
up to that radius.  Precise tuning is necessary for the pair $R_1$ and
$\delta\phi(R_1)$ to obtain a solution to all the equations.  Technically
it is easier to keep $\delta\phi(R_1)$ fixed and adjust the matching
radius
$R_1$.  If $R_1$ is chosen too small $\phi$ comes back toward, but cannot reach,  
$f_1$.  What happens is that $\phi$ eventually oscillates around $\phi=0$ as $r$ 
increases.  If $R_1$ is chosen too large $\phi$ comes back to $f_1$ at finite 
$r$ to roll over it and decreases indefinitely toward $-\infty$.  For each 
sufficiently small value of $\delta\phi$ chosen as the matching value 
there exists a desired solution with $R_1=R_c$.  The critical value $R_c$ can be 
determined numerically to arbitrary accuracy. 

\begin{figure}[htb]
\centering \leavevmode 
\mbox{
\epsfysize=6.5cm \epsfbox{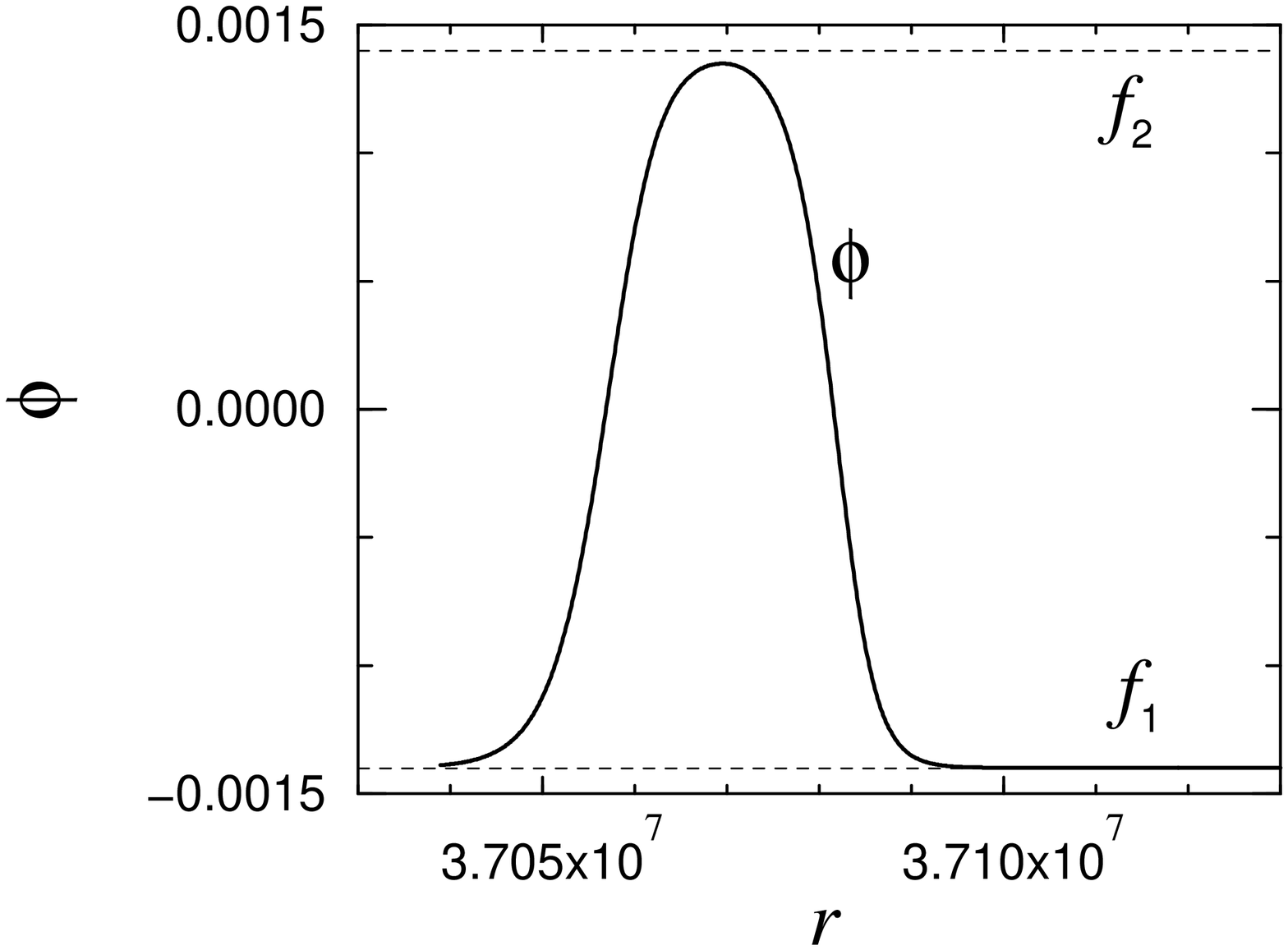}}
\caption{$\phi(r)$ of a solution with $f/m_{\rm P}= 1.40 \times 10^{-3} ,
\Delta f/f = 0.002 , \lambda = 0.01$.  $\phi$ and $r$ are in the units of
$m_{\rm P}$ and $l_{\rm P}$, respectively. The maximum value
of $\phi$ is smaller than $f_2 \sim 0.0014 \cdot m_{\rm P}$.}
\label{figphi}
\end{figure}

One example of such a solution is displayed in fig.\ \ref{figphi} for the
parameters $f/m_\P =1.40 \times 10^{-3}$, $\Delta f/f =0.002$, 
$\lambda=0.01$ and $\delta\phi(R_1)=1.0\times10^{-5}$.  The matching
radius 
$R_1$ is fine-tuned to ten digits: $R_1/l_\P \sim 3.7038855228 \times 10^7$.  
The shell is very thin compared to the radius of the shell, lying in the region
$3.704 \times 10^7 \leq r/l_\P \leq 3.710 \times 10^7$.

\begin{figure}[htb]
\centering \leavevmode 
\mbox{
\epsfysize=6.5cm \epsfbox{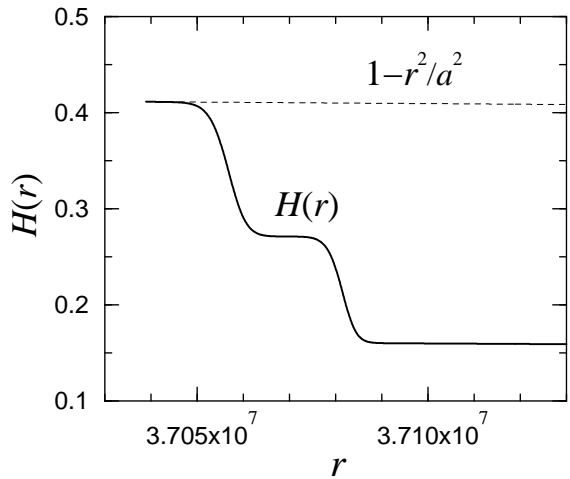}}
\caption{$H(r)$ of a solution with $f/m_{\rm P}= 1.40 \times 10^{-3},
\Delta f/f = 0.002 , \lambda = 0.01$.  $H(r)$ decreases in two steps
in the shell region.  $H(r)$ in the de Sitter space $(=1 - r^2/a^2$) is
plotted in a dotted line.}
\label{figH}
\end{figure}

\begin{figure}[htb]
\centering \leavevmode 
\mbox{
\epsfysize=6.5cm \epsfbox{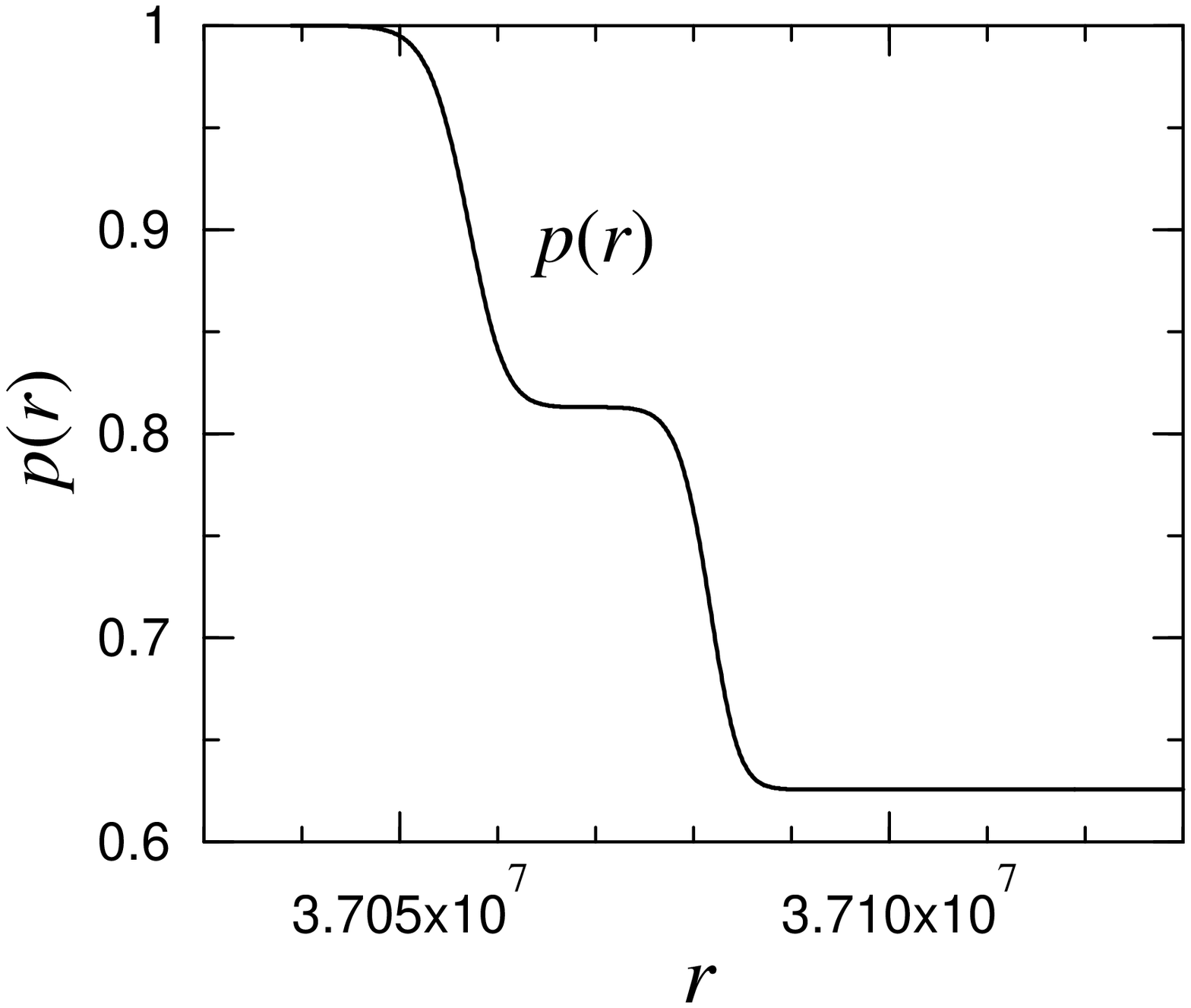}}
\caption{$p(r)$ of a solution with $f/m_{\rm P}= 1.40 \times 10^{-3},
\Delta f/f = 0.002 , \lambda = 0.01$. $p(r)$ decreases in two steps,
from 1 to 0.6256, in the shell region.}
\label{figp}
\end{figure}

In the shell region both $H(r)$ and $p(r)$ decrease in a two-step fashion. 
See figs. \ref{figH} and \ref{figp}. Inside and outside the shell $H(r)$
is given by (\ref{metric2a}) and (\ref{metric2}), whereas $p(r)$ assumes
the constant values 1 and 0.6256, respectively.

\begin{figure}[tbh]
\centering \leavevmode 
\mbox{
\epsfysize=6.5cm \epsfbox{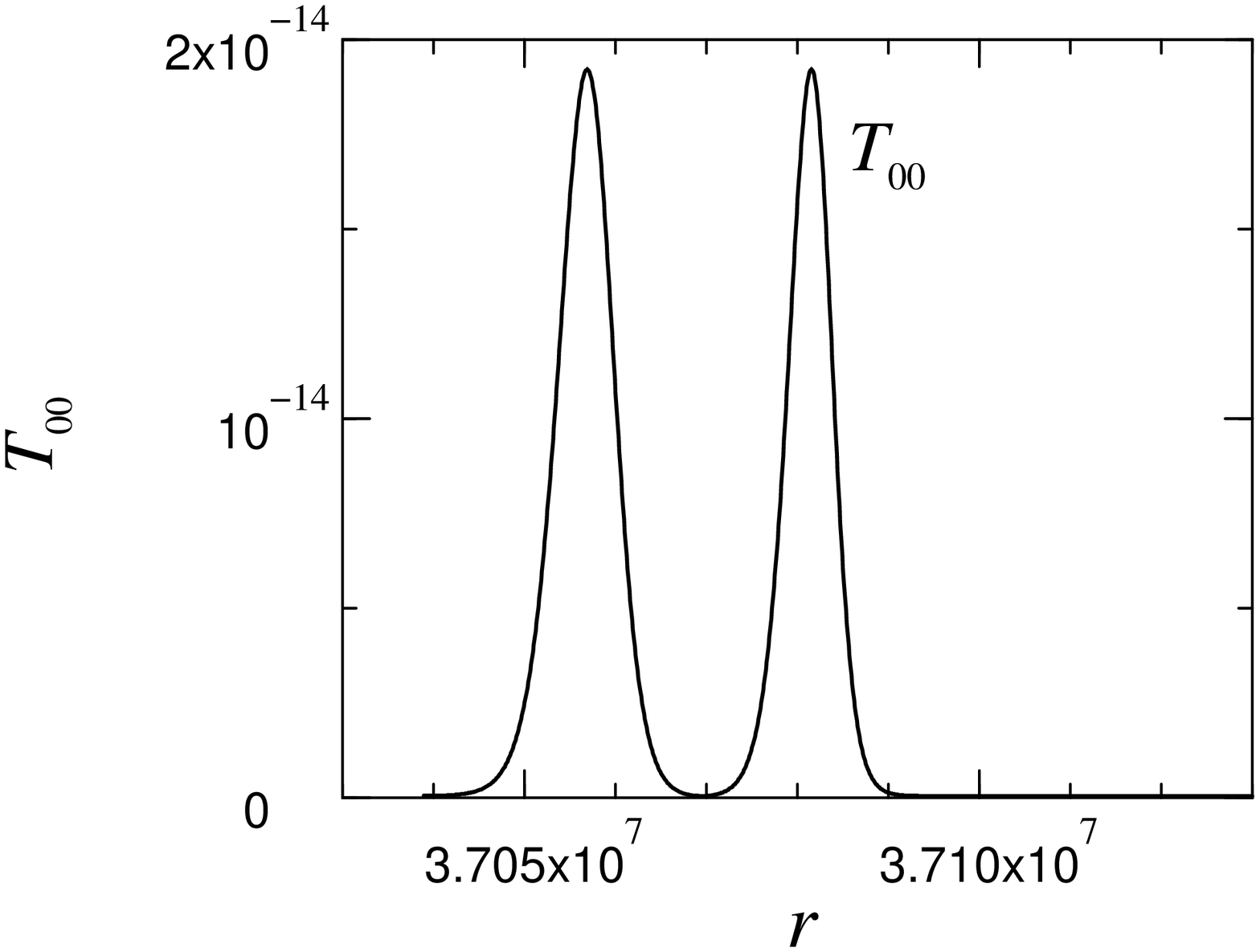}}
\caption{The energy density $T_{00}$ for a shell solution with 
$f/m_{\rm P}= 1.40 \times 10^{-3} ,
\Delta f/f = 0.002 , \lambda = 0.01$.  $T_{00}$ and $r$ are in the units
of $m_{\rm P}^4$ and $l_{\rm P}$, respectively.  The energy density is
localized in the two shells over the de Sitter background.}
\label{T00}
\end{figure}

\begin{figure}[htb]
\centering \leavevmode 
\mbox{
\epsfysize=6.5cm \epsfbox{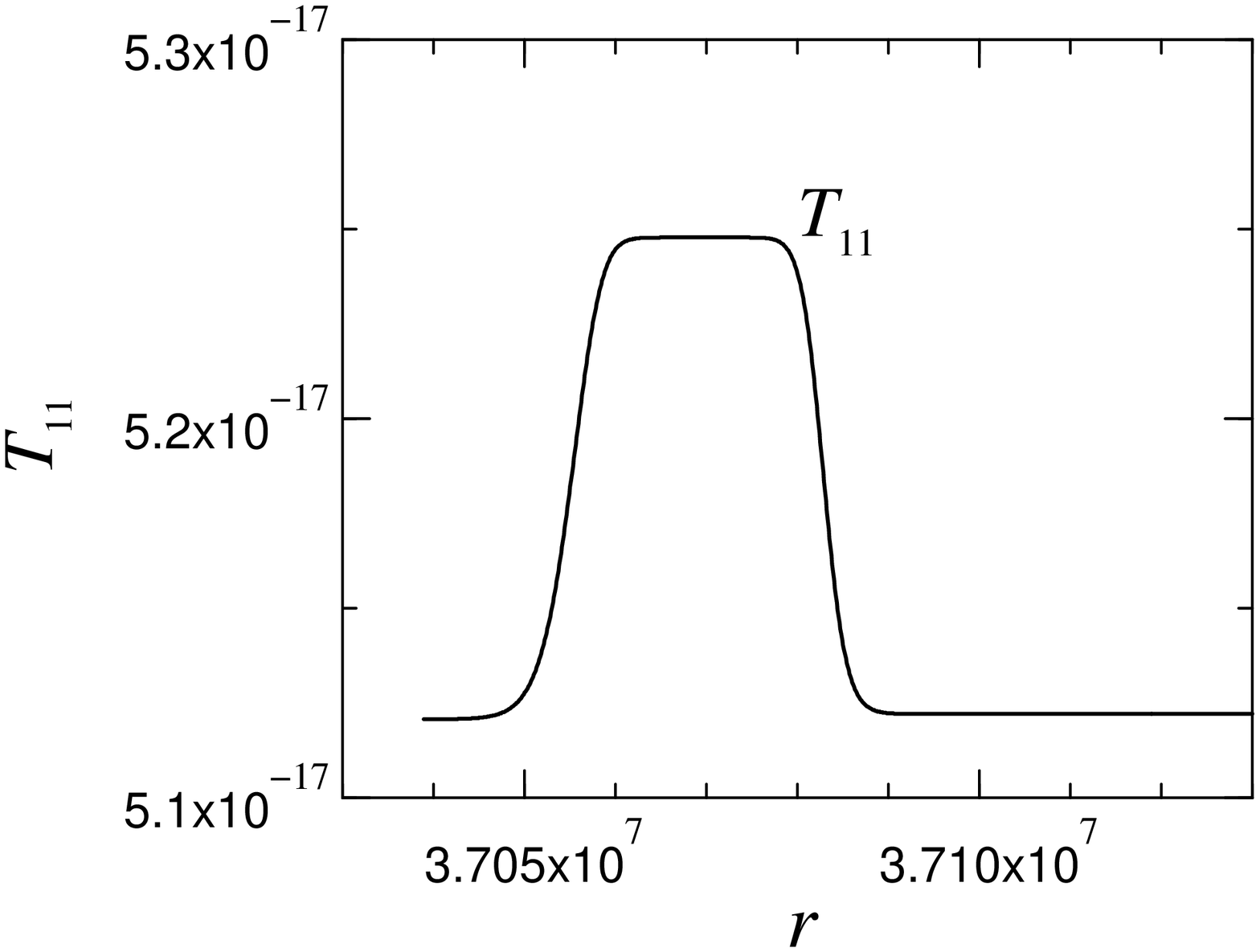}}
\caption{The radial pressure $T_{11}$ for a shell solution with 
$f/m_{\rm P}= 1.40 \times 10^{-3} ,
\Delta f/f = 0.002 , \lambda = 0.01$.  $T_{11}$ and $r$ are in the units
of $m_{\rm P}^4$ and $l_{\rm P}$, respectively.  The pressure is higher
and constant  between the two shells.}
\label{T11}
\end{figure}

The change is induced by the non-vanishing 
energy-momentum tensor components $T_{00}=- T_{22}=-T_{33}$ and $T_{11}$.
They are 
displayed in fig.\ \ref{T00} and fig.\ \ref{T11}.  The energy density 
$T_{00}$ has two sharp peaks associated with the rapid variation of $\phi$.  
The radial pressure $T_{11}$, on the other hand, steps up quickly, remains 
constant within the shell, and then steps down again.  The value of
$T_{11}$ is very small ($\sim 5 \times 10^{-17} m_{\rm P}^4$)
compared with the maximum value of 
$T_{00} (\sim 2 \times 10^{-14} m_{\rm P}^4$).  The contributions of the 
kinetic energy, $\onehalf H \phi'^2$, and potential energy, $V[\phi]$,
almost  cancel each other.

\begin{figure}[htb]
\centering \leavevmode 
\mbox{
\epsfysize=6.5cm \epsfbox{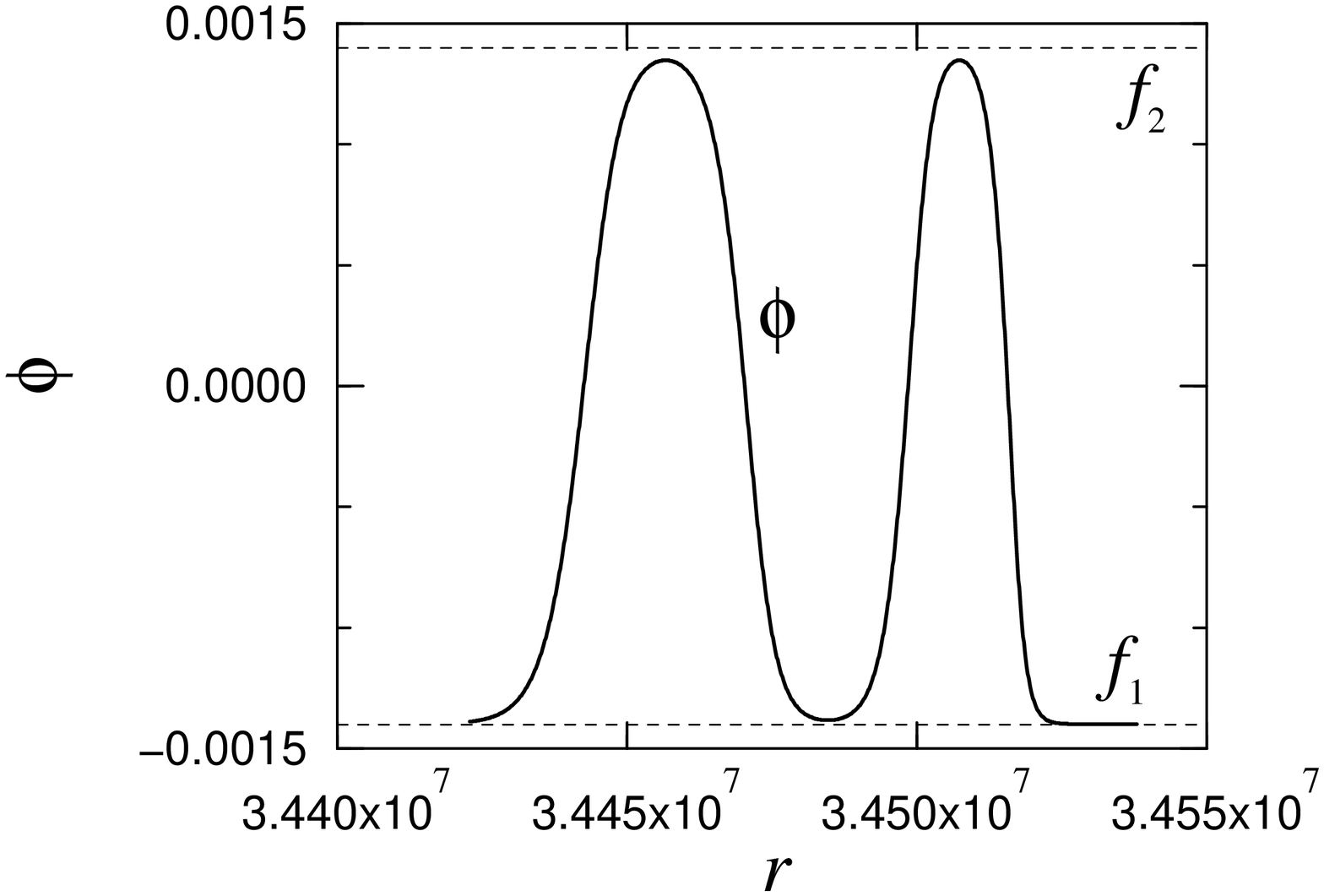}}
\caption{A solution with  double bounces  with  $f/m_{\rm P}= 1.40 \times
10^{-3} , \Delta f/f = 0.002 , \lambda = 0.01$.  A solution with multiple
bounces appears at a smaller shell radius than a solution with a single
bounce.}
\label{double}
\end{figure}

In this example $\phi(r)$ makes a bounce transition 
$f_1 \go f_2 \go f_1$ once. As the numerical value of $f$ becomes smaller
than $0.002 m_{\rm P}$ a new type of solutions with  double bounces
 $f_1 \go  f_2 \go f_1 \go f_2 \go f_1$ emerge.  One example  is  displayed
in fig.\  \ref{double}.

\begin{figure}[htb]
\centering \leavevmode 
\mbox{
\epsfxsize=8.5cm \epsfbox{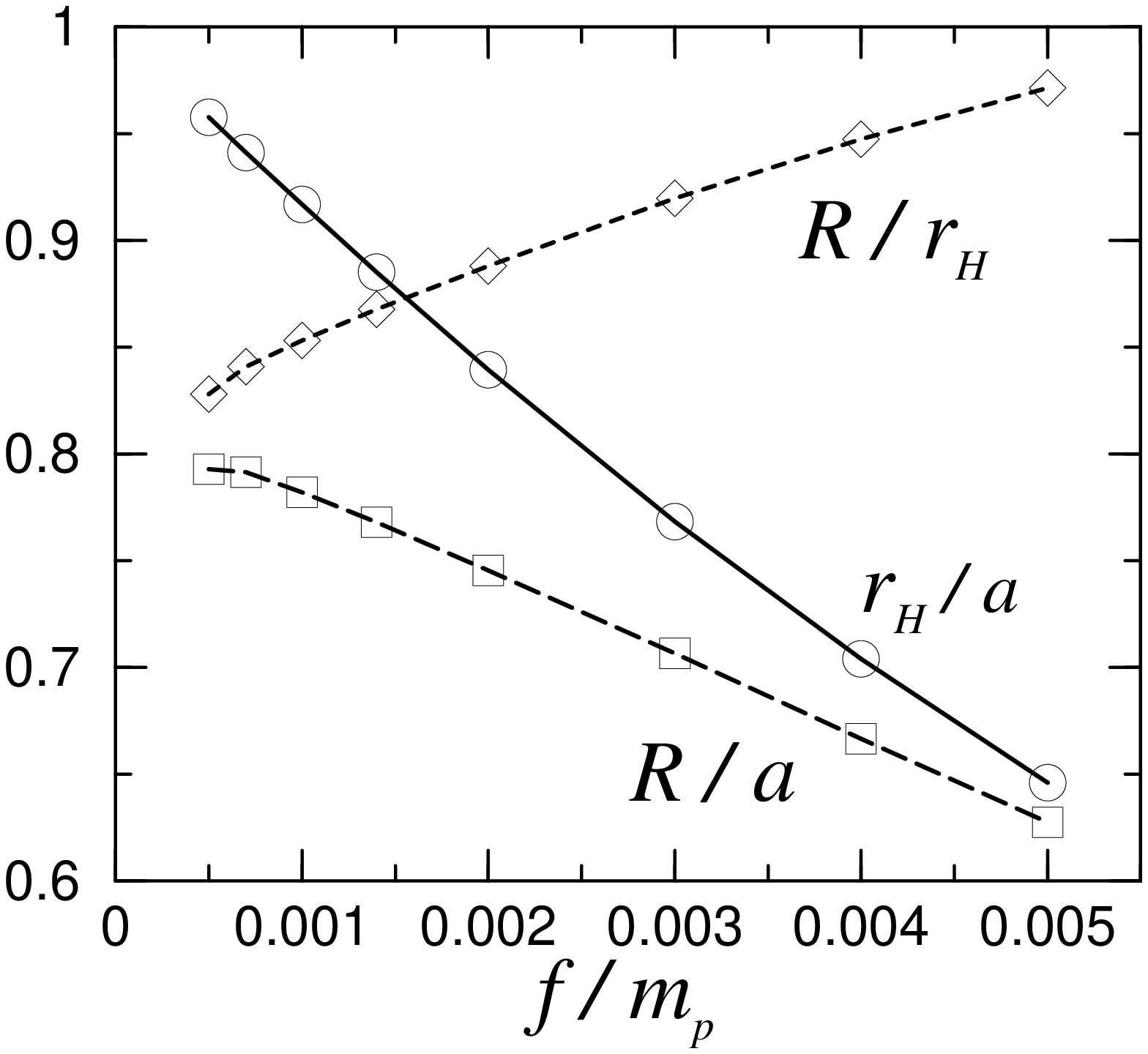}}
\caption{The $f$-dependence of the shell solutions.
$\lambda=0.01$ and $\Delta f/f=0.002$ are fixed.  $R$, $r_H$, and $a$ are
the radius of the shell, the horizon length where $H(r)$ vanishes, and
the horizon length in the de Sitter space, respectively.  The ratios of
various pairs are plotted.  Circles, squares, and diamonds correspond to
data points obtained.
}
\label{fdependence}
\end{figure}

The spectrum of the shell solutions depends on the parameter $f$.
In fig.\ \ref{fdependence} $R/a$, $r_H/a$, and $R/r_H$ are plotted as
functions of $f$.  (Finding the solution numerically becomes extremely difficult 
as $f$ becomes smaller.)  There are several features to be noted.  First,
with  fixed values of $\lambda$ and $\Delta f /f$ solutions exist only for
$f < f_{\rm max}$.  For example, with $\lambda=0.01$ and $\Delta f /f = 0.002$ 
we find that $f_{\rm max} \sim 0.006$.  Second, as $f$ goes to 0, $r_H/a \go 1$ 
and $R/a \go 0.8$.  The relative size $R/a$ of the shell remains of order
one  even in the $f \go 0$ limit.  Third, as $f$ decreases shell solutions
with multiple bounces become possible. We cannot determine how many times $\phi$ 
can bounce inside the horizon because the numerical evaluation
becomes extremely difficult when $f/m_\P < 0.001$. 

In the potential we are analysing, $\phi=f_1$ and $\phi=f_2$
correspond to the false and true vacua, respectively.  For the 
existence of  shell solutions the fact $V[f_1] > V[f_2]$ is not
crucial, however.   In the example displayed in fig.\ \ref{figphi}, for
instance,
$\phi(r)$ swings from $\phi(0)\sim f_1$ to $\phi_{\rm max} < f_2$,
and returns to $f_1$.  Numerically  $V[\phi_{\rm max}] > V[\phi(0)]$.
As we are solving classical differential equations, only the form of the
potential
$V[\phi]$ between $f_1$ and $\phi_{\rm max}$ is relevant.  The form
of the potential for $\phi > \phi_{\rm max}$ does not matter.  The
second minimum can be higher than the first one; 
$V[f_1] < V[f_2] < V[\phi_{\rm max}]$.

Before concluding the section we would like to add that we have found
no solution in which $\phi$ makes a transition from $f_1$ to $f_2$
as $r$ varies from 0 to $\infty$.

\sxn{Global Structure of the Space-Time and Solutions}

In the previous section we found novel solutions to a theory with a scalar field 
coupled to gravity in static coordinates.  The static coordinates in 
(\ref{ourmetric1}), however, do not cover all of space-time.  In this
section we  construct coordinates which allow for an extension of the
solution to the full  space-time manifold.  These coordinates are smooth
across the horizon. First we illustrate the construction with de Sitter
space, and then consider the more general case which is applied to the
shell solution.

The $R^1 \times S^3$ metric of the de Sitter space is given by
\beeq
ds^2 = a^2 \left[ -  d\tau^2 + \cosh^2 \tau 
\left( d\chi^2 + \sin^2 \chi \, d\Omega^2 \right) \right] ~.
\label{dS2}
\eneq
One may regard the de Sitter space as a hypersurface in the 
five-dimensional Minkowski spacetime constrained by the condition
\beeq
y_1^2  + y_2^2 + y_3^2 + y_4^2 = a^2 + y_0^2 ~.
\label{dS3}
\eneq
The metric (\ref{dS2}) and hypersurface (\ref{dS3}) cover the entire de Sitter
space, whereas the static metric covers only half of the space.  
Furthermore, the static metric has a coordinate singularity at $r=a$.

The relationship among these coordinate systems are easily found.
The $S^3$ metric (\ref{dS2}) can be transformed to a metric conformal to
the static Einstein universe.
\beqn
\hskip -1cm
&&
ds^2 = {a^2 \over \cos^2 \eta} 
 \left[ -  d\eta^2 
    +  d\chi^2 + \sin^2 \chi \, d\Omega^2 \right] \cr
\noalign{\kern 5pt}
\hskip -1cm
&&
\eta = {\pi\over 2} - 2 \tan^{-1} (e^{-\tau}) ~,~~ 
- {\pi\over 2} \le \eta \le {\pi\over 2}
\label{dS4}
\eeqn
Suppressing $S^2$, or $\theta$ and $\phi$ variables, one can map the whole de
Sitter space to a square region in the $\chi$-$\eta$ coordinates.  Null 
geodesics
are given by straight lines at 45 degree angles.

\begin{figure}[tbh]
\centering \leavevmode 
\rotatebox{-90}{
\epsfxsize=9.cm \epsfbox{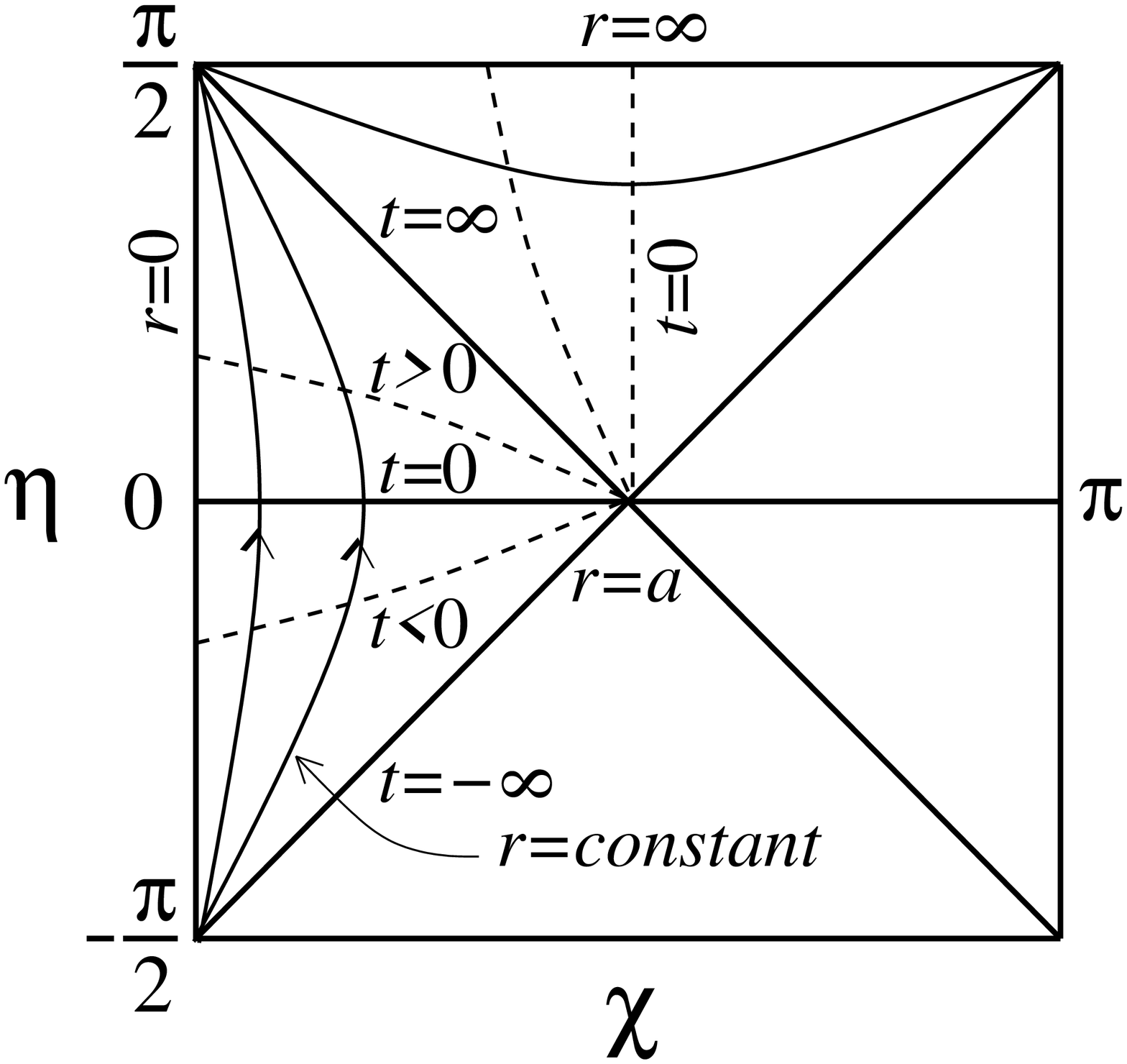}}
\caption{Penrose diagram of the de Sitter space.}
\label{Penrose1}
\end{figure}

The static and hypersurface coordinates are related by
\beqn
\hskip -1cm &&
y_0 = \cases{
\sqrt{a^2 - r^2} \, \sinh \myfrac{t}{a} &for $r<a$  \cr
\sqrt{r^2 - a^2} \, \cosh \myfrac{t}{a}  &for $r>a$   \cr}  \cr
\hskip -1cm &&
y_1 = r \cos \theta \cr
\hskip -1cm &&
y_2 = r \sin \theta \cos\phi \cr
\hskip -1cm &&
y_3 = r \sin \theta \sin\phi \cr
\hskip -1cm &&
y_4 = \cases{
\sqrt{a^2 - r^2} \, \cosh \myfrac{t}{a} &for $r<a$ \cr
\sqrt{r^2 - a^2} \, \sinh \myfrac{t}{a} &for $r>a$ \cr}
\label{connect1}
\eeqn
Similarly, (\ref{dS3}) and (\ref{dS4}) are related by
\beqn
\hskip -1cm &&
y_0 = a \tan \eta   \cr
\hskip -1cm &&
y_1 = a \sec\eta \, \sin\chi \cos \theta \cr
\hskip -1cm &&
y_2 = a \sec\eta \, \sin\chi \sin \theta \cos\phi \cr
\hskip -1cm &&
y_3 = a \sec\eta \, \sin\chi \sin \theta \sin\phi \cr
\hskip -1cm &&
y_4 = a \sec\eta \, \cos\chi \, .  
\label{connect2}
\eeqn
The region inside the cosmological horizon ($r<a$, $-\infty < t < \infty$) in 
the
static metric corresponds to the left quadrant in the conformal metric
($0 \le \chi\le \onehalf \pi$, $|\eta| < \pi/2 - \chi$) with
the relations
\beqn
{r\over a} &=& {\sin\chi\over \cos\eta} \cr
\noalign{\kern 10pt}
{t\over a} &=& {1\over 2} \ln
 {\cos\chi +\sin\eta\over \cos\chi -\sin\eta}  ~.
\label{connect3}
\eeqn
Similarly, the region outside the cosmological horizon ($r>a$, $-\infty < t <
\infty$) in the static metric corresponds to the upper quadrant in the conformal
metric ($0 \le \chi\le  \pi$, $|\eta| > \chi - \onehalf \pi$) with
the relations
\beqn
{r\over a} &=& {\sin\chi\over \cos\eta} \cr
\noalign{\kern 10pt}
{t \over a} &=& 
{1\over 2} \ln { \sin\eta + \cos\chi\over \sin\eta - \cos\chi}~.
\label{connect4}
\eeqn
See fig.\ \ref{Penrose1}.

For the general static metric given by (\ref{ourmetric1}), 
with $H=H(r)$ and 
$p=p(r)$, the construction of the conformal coordinates analogous to (\ref{dS4})
proceeds as follows.  We suppose that $H(r)$ has a single zero at $r_H$
whereas $p(r)>0$.  The new radial coordinate is defined by
\beeq
r_*(r) = \cases{
\mybig\int_0^r dr' \, \myfrac{p(r')}{H(r')} &for $r< r_H$\cr
\mybig\int_\infty^r dr' \, \myfrac{p(r')}{H(r')} &for $r> r_H$. \cr}
\label{tortoise1}
\eneq
It has a logarithmic singularity at $r_H$, diverging there as
$r_* \sim - \onehalf b\ln |r-r_H|$ where $b= -2 p(r_H)/H'(r_H) >0$.  New
coordinates are introduced by
\beqn
\tan u &=& + e^{+t/b} e^{- r_*/b} \cr
\noalign{\kern 10pt}
\tan v &=& \mp e^{-t/b} e^{- r_*/b} ~.
\label{metric3}
\eeqn
The upper sign is for $r<r_H$ and the lower sign is for $r>r_H$.
The $u$-$v$ coordinates are related to the $U$-$V$ coordinates in the
Kruskal-Szekeres (KS) coordinate system \cite{K-S} for the Schwarschild 
solution, and to the Gibbons-Hawking (GH) coordinate system \cite{G-H} for de 
Sitter space.  The connection between the KS and GH coordinate systems has 
already been discussed in \cite{Guth}.
Essentially $\tan u = U+V$ and $\tan v = V-U$.  
The metric becomes 
\beqn
ds^2 &=& - {4 b^2 F(u,v)\over \cos^2(u+v)} dudv
   + r^2 d {\Omega}^2 \cr
\noalign{\kern 10pt}
F(u,v) &=&{H\over 4 p^2} \Big( 1 - \tan u \tan v \Big)
                          \Big( 1 - \cot u \cot v \Big) \cr
\noalign{\kern 10pt}
&=& {H\over 2 p^2}  \Big( 1 \pm \cosh {2r_*\over b} \Big)
\hskip 1cm {\rm for} \left\{ 
     \matrix{r \le r_H\cr r \ge r_H\cr}  \right. ~.
\label{metric4}
\eeqn
The static metric covers the region interior to the bounding lines $u=0$,
$u+v=\onehalf \pi$, and $u - v = \onehalf \pi$. 
The horizon in the static metric, $r=r_H$,
corresponds to the single point $u=v=0$.  The function $F$ is non-vanishing and 
finite there as $e^{2r_*/b} \sim 1/|r-r_H|$.  In the
$u$-$v$ coordinates the metric is regular on $u=0$ so that the extension to the
region $u<0$ can be made, whereas the static metric covers only half of the 
space.

Applied to the de Sitter space we find
\beeq
r_H=b=a ~,~~e^{-r_*/b} = \left| {a -r\over a+ r} \right|^{1/2} ~,~~
F = 1~,
\label{connect5}
\eneq
and the metric (\ref{dS4}) is recovered by
\beeq
\eta = u + v ~,~~ \chi = v - u + {\pi\over 2}  ~.
\label{connect6}
\eneq

The shell solution found in the preceeding sections can be extended to 
the entire space-time.  In the region III ($R_2 < r $) defined  in the
static metric $H(r)$ is given by (\ref{metric2}), and 
the location of the horizon is determined by $H(r_H)=0$.  In region III,
$\delta \phi(r)$ is very small so that its equation of motion  can
be linearized. We divide region III into two;  region IIIa ($R_2 < r <
2r_H - R_2$) and  region IIIb ($r > 2 r_H - R_2$).   
In region IIIa, $H(r)$ can be approximated by
\beeq
H(r) \sim A \bigg(  1 - {r^2\over r_H^2} \bigg) ~,
\label{metric6}
\eneq
where
\beeq
A = {1\over 2} \bigg( {3r_H^2\over a^2} - 1 \bigg) ~.
\eneq
The coefficient $A$ has been chosen such that both (\ref{metric2})
and (\ref{metric6}) have the same slope $H'(r_H)$ at the horizon.  In the 
examples described in Section 3, errors caused by (\ref{metric6}) are less than 
15\% in region IIIa.  Now we write the linearized version of
(\ref{scalar2}) in  terms of $y = 1 - (r^2/r_H^2)$.
\beqn
&&\hskip -1.cm
\Bigg\{ y(1-y) {d^2\over dy^2} 
+ \bigg( 1 - {5\over 2} y \bigg) {d\over dy}
-  \bigg( \tilde \kappa^2 + {9\over 16} \bigg) \Bigg\} \, \delta\phi = 0  \cr
\noalign{\kern 5pt}
&&\hskip -1.cm
A \bigg( \tilde \kappa^2 + {9\over 16} \bigg)
 = \kappa^2 + {9\over 16} = {1\over 4} \omega^2 a^2 
\label{scalar5}
\eeqn
Since $\delta\phi(r)$ must be regular at $r=r_H$ ($y=0$), the solution is
\beeq
\delta\phi(r) = \delta\phi(r_H) \cdot
F( \hbox{$\frac{3}{4}$} + i\tilde\kappa, \hbox{$\frac{3}{4}$} - i\tilde\kappa,
 1; y)  ~.
\label{phi3}
\eneq
The normalization $\delta\phi(r_H)$ must be such that $\delta\phi(r)$ matches at
$r=R_2$ with the value determined by numerical integration in region II.
Essentially $\delta\phi(r)$ decreases exponentially when $R_2 < r < r_H$.

Near $r=R_2$ the hypergeometric function behaves as 
\beeq
F( \hbox{$\frac{3}{4}$} + i\tilde\kappa, 
   \hbox{$\frac{3}{4}$} - i\tilde\kappa, 1; y)
\sim {1\over 2 \sqrt{\pi \tilde\kappa} } \,
   y^{-1/4} (1-y)^{-1/2} \, 
  \exp \Big\{ 2 \tilde\kappa \sin^{-1} \sqrt{y} \Big\} 
\label{geometricF3}
\eneq
so that
\beeq
{\delta \phi'(r) \over \delta \phi(r)} 
= - {2r\over r_H^2} \cdot {\tilde\kappa\over \sqrt{y(1-y)} }  ~.
\label{phi4}
\eneq
The value of the right side of (\ref{phi4}) at $r=R_2$ can be compared with the
value obtained by direct numerical integration in region II.  In one example
with $f/m_{\rm P}= 2\times 10^{-3}, \Delta f/f= 2\times 10^{-3}, 
\lambda= 0.01$, $R_2 = 1.765 \times 10^7, a = 2.365 \times 10^7, 
r_H = 1.985 \times 10^7$, $\kappa = 3344$, and $\tilde\kappa = 4481$.
The numerical value for $\delta \phi'(R_2) / \delta \phi(R_2)$ is
$-0.000848$, whereas the value from (\ref{phi4}) is $-0.000985$.  
With the uncertainty in the value of $\tilde\kappa$ caused by the
approximation (\ref{metric6}) taken into account, one may conclude that the
agreement is rather good.  To gauge the difficulty of determining the
solution numerically for all values of $r$, we note that 
$\delta \phi(r_H) \sim 10^{-1850} \cdot \delta\phi(R_2)$.

The most important observation is that the solution $\phi(r)$ is regular at
the horizon, $r=r_H$, though the static metric $(t,r,\theta,\phi)$ is not:
$r=r_H$ in the static metric is a coordinate singularity.  As seen above,
$r=r_H$ corresponds to $u=v=0$ in the conformal metric (\ref{metric4})
which is non-singular.   Near the horizon
$\tan u \tan v \sim r - r_H$ so that
\beeq
\phi = f_1 + \delta \phi(r_H) 
\bigg\{ 1 - {2\over r_H} \Big( \tilde \kappa^2 + {9\over 16} \Big)
  \, uv + \cdots  \bigg\} ~.
\label{phi5}
\eneq
Furthermore, as $\tan u \tan v = e^{-2 r_*(r)/b}$, the static solution
$\phi(u,v)$ depends only on $\tan u \tan v$ in the entire space-time.  This 
implies that
\beeq
\phi(-u, -v) = \phi(u, v) ~.
\label{phi6}
\eneq
In the $\eta$-$\chi$ coordinates ($S^3$-metric) defined by (\ref{connect6})
\beeq
\phi(\eta, \chi) = \phi(-\eta, \pi - \chi) = \phi(\eta, \pi-\chi) ~.
\label{phi7}
\eneq
This relation shows that there is a mirror image of the shell structure
in the other hemisphere in the $S^3$-metric.  See fig.\ 12.  The shaded
regions in the figure represent the shells.  The size of the shells 
is invariant, as is obvious in the static coordinates.  In the 
$S^3$ or conformal metric an observer at the center ($\chi = 0 $ or $\pi$)
sends a light signal toward the shell.  Another observer at the shell,
upon receiving the signal from the center, sent a signal back to
the observer at the center.  The total elapsed time, in terms of the
proper time of the observer at the center $a\tau$ where $\tau$
is related to $\eta$ by (\ref{dS4}) at $\chi=0$, is 
$a \ln (a+R)/(a-R)$, independent of when the signal is emitted  
initially, as straightforward manipulations show.

\begin{figure}[tbh]
\centering \leavevmode 
\rotatebox{-90}{
\epsfxsize=9.cm \epsfbox{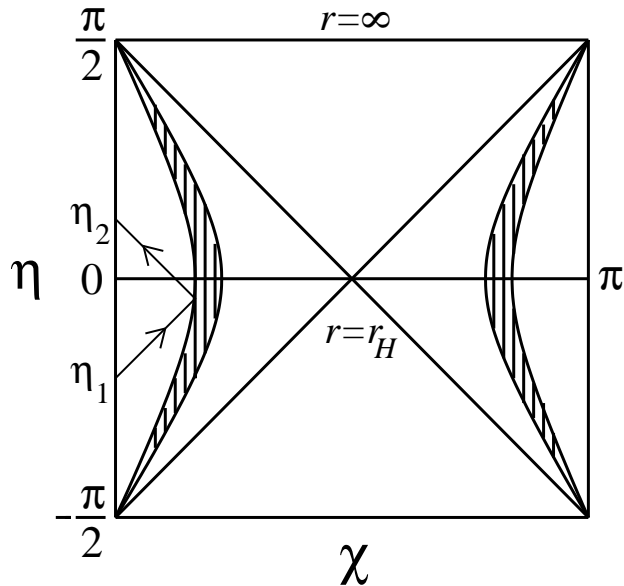}}
\caption{Global structure of the shell solution in the conformal coordinates.
The mirror image appears in the other hemisphere.}
\label{Penrose2}
\end{figure}

Finally, in the region outside the horizon in the static metric, or
in the upper or lower quadrant in the Penrose diagram,  $\delta\phi(r)$
decays as $r^{-3/2}$.    The proof goes as follows.
For $r \gg r_H$,
$H \sim 1 - (r^2/a^2)$, so that $\delta\phi(r)$ satisfies Eq.\ 
(\ref{scalar3}).  Hence it  is a linear combination of 
$u_1 = F(\frac{3}{4} + i \kappa , \frac{3}{4} - i \kappa ,\frac{3}{2} ; z)$
and $u_2 = z^{-1/2} 
F(\frac{1}{4} + i \kappa , \frac{1}{4} - i \kappa ,\frac{1}{2} ; z)$.
These can be written as
\beqn
&&\hskip -1cm
\left[ \matrix{u_1 \cr u_2\cr} \right] =
\left[ \matrix{\onehalf \Gamma( \hbox{${3\over 4}$} - i\kappa )^{-2} \cr
    \Gamma( \hbox{${1\over 4}$} - i\kappa )^{-2} \cr} \right]
{\sqrt{\pi} \,  \Gamma (-2i\kappa) \over (-z)^{(3/4) + i\kappa}}
~ F \Big( \hbox{$\frac{3}{4}$} + i\kappa, 
   \hbox{$\frac{1}{4}$} + i\kappa, 1 + 2i\kappa ; {1\over z} \Big)  \cr
\noalign{\kern 12pt}
&&\hskip .1cm
+ \left[ \matrix{\onehalf \Gamma( \hbox{${3\over 4}$} + i\kappa )^{-2} \cr
    \Gamma( \hbox{${1\over 4}$} + i\kappa )^{-2} \cr} \right]
{\sqrt{\pi} \,  \Gamma (+ 2i\kappa) \over (-z)^{(3/4) - i\kappa}}
~ F \Big( \hbox{$\frac{3}{4}$} - i\kappa, 
   \hbox{$\frac{1}{4}$} - i\kappa, 1 - 2i\kappa ; {1\over z} \Big) ~.  
\label{geometricG4}
\eeqn
From this one can write
\beeq
\phi \sim f_1 + A \left( {a\over r} \right)^{3/2} \,
\sin \Big( 2\kappa \ln {r\over a} + \delta \Big)
\label{phi8}
\eneq
when $r \gg r_H$.

\sxn{False Vacuum Decay}

If the region inside the shell is in a false vacuum state one should consider 
its quantum decay to the true vacuum state.
The lifetime of the false vacuum may be determined semi-classically using the
methods of Coleman {\it et al.} without \cite{without} or with \cite{Coleman}
gravity taken into account.  The rate per unit volume for making a transition
from the false vacuum to the true vacuum is expressed as
\begin{equation}
\frac{\Gamma}{V} = Ae^{-B/\hbar}[1+{\cal O}(\hbar)]
\end{equation}
where Planck's constant has been used here to emphasize the semi-classical 
nature of the tunneling rate.  For the potential being used in this paper
we find the O(4), Euclidean space, bounce action (neglecting gravity) to be
\begin{equation}
B_0 = \frac{36\pi^2}{\lambda}\left(\frac{f}{\Delta f}\right)^3
\end{equation}
where the radius of the critical size bubble which nucleates the transition is
\begin{equation}
\rho_c=\frac{3}{\Delta f} \sqrt{\frac{2}{\lambda}} \, .
\end{equation}
For any sensible estimate of the coefficient $A$ the lifetime of the false
vacuum will exceed the present age of the universe when the condition
\begin{equation}
\lambda \left(\frac{\Delta f}{f}\right)^3 < 1
\end{equation}
is satisfied.  This calculation is based on the thin wall approximation, which 
is valid when the critical radius is large compared to the coherence length of 
the potential, namely $1/\sqrt{|V^{\prime\prime}|}$.  This condition translates 
into
\begin{equation}
\Delta f \ll 6f \, .
\end{equation}
With gravity included the bounce action is
\begin{equation}
B = \frac{B_0}{\left[1+(\rho_c/2R)^2 \right]^2} \, .
\end{equation}
Gravitational effects are negligible when $\rho_c < R$.  When this condition
is not fulfilled the shell radius is too small to accommodate even a single 
nucleation bubble and therefore nucleation is further suppressed.

\sxn{Summary}

In this paper we have reported the discovery of shell-like solutions to the 
combined field equations of gravity and a scalar field with a double-minima 
potential.  These solutions exist in a space that is asymptotically de Sitter.
The range of parameters which allow such solutions are very broad.  If anything 
like these structures exist in nature they most likely would have been created 
in the early universe and are therefore cosmological.  We know of no other way 
to produce them.

To make matters interesting, let us suppose that the cosmological constant
suggested by recent observations of distant Type Ia supernovae \cite{super}
arises from the universe actually being in a false vacuum state.  A best fit to
all cosmological data \cite{science} reveals that the present energy density of
the universe has the critical value of $\epsilon_c = 3H_0^2/8\pi G$,
with one-third of it consisting of ordinary matter and two-thirds of it
contributed by the cosmological constant.  Suppose that the cosmological
constant arises from a potential of the form we have been analyzing.
Then with a present value of the
Hubble constant of $H_0=65$ km/s$\cdot$Mpc we find that
\begin{equation}
a = \sqrt{\frac{3}{2}}\frac{1}{H_0} = 1.7 \times 10^{26} \,\, {\rm m}
\end{equation}
and so
\begin{equation}
\left( \lambda f^3 \Delta f \right)^{1/4}=2.4 \times 10^{-3} \,\, {\rm eV} \, .
\end{equation}
This is a constraint on the parameters of the potential, $\lambda$, $f$ and 
$\Delta f$.  Although they cannot be determined individually from this data,
we can place limits on them such that a shell structure might arise.  Recalling
(\ref{estimate2}) we find
\begin{eqnarray}
f &<& 30 \lambda^{-1/6} \, {\rm MeV} ~, \nonumber \\
w &>& 6.3 \lambda^{-1/3} \, {\rm fm} ~.
\end{eqnarray}
This is pure speculation of course.  A cosmological constant, if
it exists, may have its origins elsewhere.  But if it does arise from a false
vacuum, a variety of questions immediately present themselves.  Is $\phi$ a new
field, not present in the standard model of particle physics, whose only purpose
is this?  Where does the energy scale of 2.4 meV come from?  Why should
$V[\phi]$ have a global minimum of 0, especially when quantum mechanical
fluctuations are taken into account?  To these questions we have no answers.

\vskip 1cm

\leftline{\bf Acknowledgments}

This work was supported by the US Department of Energy under Grant
DE-FG02-87ER40328, and by the Japan Ministry of Education and Science 
under Grants No.\ 13640284 and No.\ 13135215.

\vskip .5cm

\leftline{\bf References}  

\renewenvironment{thebibliography}[1]
        {\begin{list}{[$\,$\arabic{enumi}$\,$]}  
        {\usecounter{enumi}\setlength{\parsep}{0pt}
         \setlength{\itemsep}{0pt}  \renewcommand{\baselinestretch}{1.2}
         \settowidth
        {\labelwidth}{#1 ~ ~}\sloppy}}{\end{list}}

\end{document}